\documentclass[a4paper,11pt]{article}
\pdfoutput=1 % if your are submitting a pdflatex (i.e. if you have
\usepackage{jheppub} % for details on the use of the package, please
\usepackage[T1]{fontenc} % if needed
\newcommand{\hs}{\hspace*{0.5cm}}

\newcommand{\be}{\begin{equation}}
\newcommand{\ee}{\end{equation}}
\newcommand{\bea}{\begin{eqnarray}}
\newcommand{\eea}{\end{eqnarray}}
\newcommand{\ben}{\begin{enumerate}}
\newcommand{\een}{\end{enumerate}}
\newcommand{\bde}{\begin{widetext}}
\newcommand{\ede}{\end{widetext}}

\newcommand{\crn}{\nonumber \\}

\newcommand{\al}{\alpha}

\newcommand{\bet}{\beta}

\newcommand{\om}{\omega}

\newcommand{\+}{\dagger}
\newcommand{\fr}{\frac}
\newcommand{\bc}{\begin{center}}
\newcommand{\ec}{\end{center}}

\title{\boldmath Comment on "Flavored leptogenesis and neutrino mass with $A_4$
symmetry", JHEP12(2021)051}
\author[a,b,1]{V. V. Vien,\note{Corresponding author.}}
\affiliation[a]{Institute of Applied Technology, Thu Dau Mot University, Binh Duong Province, Vietnam}
\affiliation[b]{Department of Physics, Tay Nguyen University, 567 Le Duan street, Buon Ma Thuot city,
Daklak province, Vietnam.}
\emailAdd{vovanvien@tdmu.edu.vn}
\abstract{Recently, in ref. \cite{A4Datta} Datta et al. proposed an $A_4$ flavor symmetric model supplemented by $Z_2\times Z_3$ symmetry which can accommodate the appropriate lepton mixing and neutrino masses via Type-I seesaw mechanism. They
have constructed a minimal  model with only one $SU(2)_L$ doublet scalar and six flavons that generate a specific flavor
structure, favors the normal hierarchy of light neutrinos and narrows down
the range of Dirac CP violating phase. Taking into account the contribution of all invariant terms, under all symmetries, is very important in the model building process, however, in ref. \cite{A4Datta} the authors miss a Majorana mass term which contributes to and changes the structure of the Majorana mass matrix of the right handed neutrinos. In this comment paper, we point out the above issue and provide a solution to fill in the missing term in ref. \cite{A4Datta}.}

\begin{document}
\maketitle
\flushbottom
\section{Comments}
A recent study with $A_4$ symmetry\footnote{The models in refs. \cite{A4Dattaconf} and \cite{A4Datta} own the same particle and scalar contents.} made by Datta et al. \cite{A4Datta}, with the given particle and scalar contents, the Yukawa term\footnote{In fact, under $S_4$, $\overline{N^c_R} N_R$ contains two invariant terms $(\overline{N^c_R} N_R)_{\mathbf{3_s}}$ and $(\overline{N^c_R} N_R)_{\mathbf{3_a}}$, however, $(\overline{N^c_R} N_R)_{\mathbf{3_a}}\sim \mathbf{3_a}(\overline{N^c_{2R}} N_{3R}-\overline{N^c_{3R}} N_{2R}, \overline{N^c_{3R}} N_{1R}-\overline{N^c_{1R}} N_{3R}, \overline{N^c_{1R}} N_{2R}-\overline{N^c_{2R}} N_{1R})$ vanishes due to the
symmetric in $N_{iR}$ and $N_{jR}$.} $\frac{z^\nu}{2\Lambda} \left[\left(\overline{N^c_R} N_R\right)_{\mathbf{3_s}} \left(\varphi \Psi^* +\varphi^* \Psi\right)_{\mathbf{3}}\right]_{\mathbf{1}}$ must exist %in the neutrino sector since
because they are invariant under all symmetries of the model $G_{\mathrm{SM}}\times A_4\times Z_2\times Z_3$. Therefore, the corresponding interaction term has to be added to the Lagrangian expression in eq. (2.1) of ref. \cite{A4Datta} for a more complete contribution and the Majorana mass matrix $M_R$ becomes:
\begin{eqnarray}\label{mat:heavy}
 M_{R} =\left(
\begin{array}{ccc}
 M & 0   & f_R \\
 0  & M & 0\\
 f_R & 0 & M
\end{array}
\right), %\hs \Blue{f_R = 2 z^\nu \frac{v_{\varphi} v_{\Psi}}{\Lambda}}.
\label{MRv}
\end{eqnarray}
with  \bea
f_R &=&2 z^\nu \frac{v_{\varphi} v_{\Psi}}{\Lambda}. \label{fRv} \eea
In the diagonal charged lepton basis, the light neutrino mass matrix takes the form:
  \bea
 m_{\nu}&=&-v^2 V^\+ \mathbf{Y}^\nu M_R^{-1} \mathbf{Y}^{\nu T} V^* =-\frac{1}{M}  V^\+ \left(v^2 \mathbf{Y}^\nu \mathbf{Y}^{\nu T}\right) V^*,  \label{mnu} \eea
 where
 \bea && \mathbf{Y}^\nu \mathbf{Y}^{\nu T}=Y^\nu Y^{\nu T}\crn
 &&+ \left(
\begin{array}{ccc}
 \frac{f_R \left\{2 \left(f_1^\nu-f_2^\nu\right) f_3^\nu M-\left[(f_1^\nu-f_2^\nu)^2+f_3^{\nu^2}\right] f_R\right\}}{(f_R-M) (f_R+M)} & 0 & \frac{f_R \left[\left(f_1^{\nu^2}-f_2^{\nu^2}+f_3^{\nu^2}\right) M-2 f_1^{\nu} f_3^{\nu} f_R\right]}{(f_R-M) (f_R+M)} \\
 0 & 0 & 0 \\
 \frac{f_R \left[\left(f_1^{\nu^2}-f_2^{\nu^2}+f_3^{\nu^2}\right) M-2 f_1^{\nu} f_3^{\nu} f_R\right]}{(f_R-M) (f_R+M)} & 0 & \frac{f_R \left\{2 \left(f_1^{\nu}+f_2^{\nu}\right) f_3^{\nu} M-\left[\left(f_1^{\nu}+f_2^{\nu}\right)^2+f_1^{\nu^2}\right] f_R\right\}}{(f_R-M)(f_R+M)} \\
\end{array}
\right),\hs \eea
which is different from that of ref. \cite{A4Datta}, $Y^\nu Y^{\nu T}$. Here, $f_R$ is defined in eq. (1) while $V$ and $Y^\nu Y^{\nu T}$ are, respectively, given by eqs. (2.9) and (2.14) of ref. \cite{A4Datta} and all the other parameters ($M$ and $f^\nu_{1,2,3}$) are the same as those of ref. \cite{A4Datta}.

We now rewrite $m_{\nu}$ in eq. (\ref{mnu}) in the following form: \bea
 m_{\nu}&=&V^\+ X_\nu  V^*, \label{mnuv} \eea
 where
\bea
&& X_\nu=-\frac{v^2}{M} \mathbf{Y}^\nu \mathbf{Y}^{\nu T}=\left(
\begin{array}{ccc}
 A_1 & 0 & B \\
 0 & A_2 & 0 \\
 B & 0 & A_3 \\
\end{array}
\right), \label{mnu1}
   \eea
with
\bea
&&A_1=\frac{\left\{ \left[\left(f_1^\nu - f_2^\nu\right)^2 + f_3^{\nu^2}\right] M + 2 \left(f_2^\nu-f_1^\nu\right) f_3^\nu f_R\right\} v^2}{f_R^2-M^2}\equiv A_{01} e^{i\al_1},\crn
&&A_3=\frac{\left\{ \left[\left(f_1^{\nu}+f_2^{\nu}\right)^2+f_3^{\nu^2}\right]M-2  \left(f_1^{\nu}+f_2^{\nu}\right) f_3^{\nu} f_R\right\} v^2}{f_R^2-M^2}\equiv A_{03} e^{i\al_3},\crn
&&A_2=-\frac{v^2 f_3^{\nu^2}}{M}\equiv A_{02} e^{i\al_2},\, B=\frac{\left[2 f_1^\nu f_3^\nu M-\left(f_1^{\nu^2} - f_2^{\nu^2} + f_3^{\nu^2}\right) f_R\right] v^2}{f_R^2-M^2}\equiv B_{0} e^{i\beta}.
\eea
In order to diagonalise $m_{\nu}$ in eq. (\ref{mnuv}), we define a Hermitian matrix, given by
\bea
\mathrm{M}^2_\nu&=& m^{+}_{\nu} m_{\nu}=V X^2_\nu V^*,\label{Mnusq}
\eea
where
\bea
X^2_\nu&=& X^{+}_{\nu} X_{\nu}
= \left(
\begin{array}{ccc}
A_0^2 & 0 & D_0^2 e^{-i \psi} \\
 0 & B_0^2 & 0 \\
D_0^2 e^{i \psi} & 0 & C_0^2 \\
\end{array}
\right),\label{Xnu}
\eea
with
\bea
&&A^2_0 = A_{01}^2 + B_0^2, \hs  B_0^2 = A_{02}^2,\hs C_0^2 = A_{03}^2 + B_0^2, \crn
&&D_0^2 = B_0\sqrt{A_{01}^2 + A_{03}^2 + 2 A_{01} A_{03} \cos(\al_1 + \al_3 - 2 \bet)}, \label{ABCD0sq}\\
&&\psi=\arcsin\left(\frac{A_{01} \sin (\al_1-\beta)-A_{03} \sin (\al_3-\beta)}{\sqrt{A_{01}^2+A_{03}^2+2 A_{01} A_{03} \cos (\al_1+\al_3-2 \beta)}}\right). \label{psi}
\eea
The mass matrix $X_\nu$ in eq. (\ref{Xnu}) is diagonalised by the unitary matrix $U_{13}$, satisfying
 \bea
U_{13}^\+ X^2_\nu U_{13}=\left\{
\begin{array}{l}
\left(
\begin{array}{ccc}
m^2_1 & 0 & 0 \\
0 & m^2_{2} & 0 \\
0 & 0 & m^2_{3}%
\end{array}%
\right),\hspace{0.2cm} U_{13}=\left(
\begin{array}{ccc}
\cos\theta & 0 & \sin \theta. e^{-i \psi}\\
 0 & 1 & 0 \\
 -\sin \theta. e^{i \psi} & 0 &\cos\theta  \\
\end{array}
\right) \hspace{0.2cm}\mbox{for NH,}\ \  \\
\left(
\begin{array}{ccc}
m^2_{3} & 0 & 0 \\
0 & m^2_{2} & 0 \\
0 & 0 & m^2_1
\end{array}%
\right) ,\hspace{0.2cm} U_{13}=\left(
\begin{array}{ccc}
 \sin \theta & 0 &  -\cos\theta. e^{-i \psi}\\
 0 & 1 & 0 \\
 \cos\theta . e^{i \psi} & 0 & \sin \theta \\
\end{array}
\right) \hspace{0.2cm}\mbox{for IH, }%
\end{array}%
\right.  \label{Unuv}
\eea
where
\bea
&&m^2_{1,3} =\frac{1}{2} \left(A_0^2 + C_0^2 \mp \sqrt{\left(A_0^2 - C_0^2\right)^2 + 4 D_0^4}\right)\equiv X^2_0\mp Y^2_0, \,\,\, m^2_2=B_0^2, \label{m123sq}\\
%&&\psi=\arcsin\left(\frac{A_{01} \sin (\al_1-\beta)-A_{03} \sin (\al_3-\beta)}{\sqrt{A_{01}^2+A_{03}^2+2 A_{01} A_{03} \cos (\al_1+\al_3-2 \beta)}}\right), \label{psi}\\
%&&\theta=%\arctan\left(\frac{m^2_3 -C_0^2}{D_0^2}\right)=\arctan\left(\frac{A^2_0-m^2_1}{D_0^2}\right). %\arccos\left(\frac{1}{\sqrt{\frac{\left(\sqrt{\left(\text{A01}^2-\text{A03}^2\right)^2+4 \text{D0}^4}+\text{A01}^2-\text{A03}^2\right)^2}{4 \text{D0}^4}+1}}\right)\crn
%&&=
&&\theta=\arccos\left[\left(1+\frac{\left(A_{0}^2-C_{0}^2+\sqrt{\left(A_{0}^2-C_{0}^2\right)^2+4 D_0^4}\right)^2}{4 D_0^4}\right)^{\hspace{-0.15 cm}-\frac{1}{2}}\right].  \label{theta}
%\arccos\left(\frac{\left(\sqrt{\left(A_{01}^2-A_{03}^2\right)^2+4 D_0^4}+A_{01}^2-A_{03}^2\right)^2}{4 D_0^4}+1\right)^{-1/2}.
\eea
%The sum of neutrino mass is given by:
%\bea
%\sum m_\nu =\beta_0+\sqrt{\beta_0^2-\Delta m^2_{21}}+\sqrt{\beta_0^2-\Delta m^2_{21}+\Delta m^2_{31}}. \label{sum}
%\eea
Combining eqs. (\ref{Mnusq}) and (\ref{Unuv}) yields
\bea
\left(V U_{13}\right)^\+ M^2_\nu \left(V U_{13}\right)=\left\{
\begin{array}{l}
\mathrm{diag}\left(m^2_1,\, m^2_{2},\, m^2_{3}\right) \hspace{0.2cm}\mbox{for NH,}\ \  \\
\mathrm{diag}\left(m^2_3,\, m^2_{2},\, m^2_{1}\right) \hspace{0.2cm}\mbox{for IH,}
\end{array}%
\right.  \label{Unudiag}
\eea
i.e., the leptonic mixing matrix, determined by $U_{\mathrm{Lep}}=V U_{13}$, has the form:
\bea
U_{\mathrm{Lep}}=\left\{
\begin{array}{l}
\fr{1}{\sqrt{3}}\left( \begin{array}{ccc}
 \cos\theta-\sin\theta. e^{i \psi }      &1 & \cos\theta + \sin\theta. e^{-i \psi }  \\
\cos\theta - \om^2 \sin\theta. e^{i \psi} &\om & \om^2 \cos\theta +  \sin\theta. e^{-i \psi}  \\
\cos\theta - \om \sin\theta. e^{i \psi}&\om^2  & \om \cos\theta +  \sin\theta. e^{-i \psi} \\
\end{array}\right) \hspace{0.1cm}\mbox{for NH},  \label{Ulepton}  \\
\fr{1}{\sqrt{3}} \left( \begin{array}{ccc}
 \sin\theta + \cos\theta. e^{i \psi }      &1 & \sin\theta - \cos\theta. e^{-i \psi}  \\
\sin\theta + \om^2 \cos\theta. e^{i \psi} &\om & \om^2 \sin\theta - \cos\theta. e^{-i \psi}  \\
 \sin\theta+ \om \cos\theta. e^{i \psi}&\om^2  & \om \sin\theta - \cos\theta. e^{-i \psi} \\
\end{array}\right) \hspace{0.1cm}\mbox{for IH}.
\end{array}%
\right.
\eea
Expression (\ref{Ulepton}) implies $U_{\mathrm{Lep}}$ possesses the $\mathrm{TM}_2$ form sincce $|\left(U_{\mathrm{Lep}}\right)_{i2}|=\frac{1}{\sqrt{3}}\, (i=1,2,3)$.

Now, comparing eq. (\ref{Ulepton}) with the standard parameterization of the leptonic mixing matrix\footnote{Here, $c_{ij}=\cos \theta_{ij}$, $s_{ij}=\sin \theta_{ij}$ with
$\theta_{12}$, $\theta_{23}$ and $\theta_{13}$ are the solar
angle, atmospheric angle and the reactor angle, respectively; $\delta_{CP}$ is the Dirac phase and $\eta_{1, 2}$ are two Majorana phases.} \footnote{In general, %$\mathrm{P}$ takes the form
$\mathrm{P}=\mathrm{diag}\left(e^{i\omega_{1}}, e^{i\omega_{2}}, e^{i\omega_{3}}\right)$ where $\omega_{i} \, (i=1,2,3)$ are three Majorana phases which are specified by two combinations %made of $\omega_{1,2,3}$
such as $\omega_{i}-\omega_{1}$, $\omega_{i}-\omega_{2}$ and $\omega_{i}-\omega_{3}$ in place of $\omega_{i}$ %$\, (i=1, 2, 3)$
in $U_{\mathrm{PMNS}}$ matrix. In current work, $\omega_i$ is specified by two combinations $\omega_{i}-\omega_{2}$; thus, $\mathrm{P}=\mathrm{diag}\left(e^{i(\omega_{1}-\omega_{2})}, 1, e^{i(\omega_{3}-\omega_{2})}\right)\equiv\mathrm{diag}\left(e^{i\eta_{1}}, 1, e^{i\eta_{2}}\right)$.} $\mathrm{U_{PMNS}}$,
\bea
       \mathrm{U}_{\mathrm{PMNS}} =\left(
\begin{array}{ccc}
c_{12} c_{13} & c_{13} s_{12} & e^{-i \delta_{CP}} s_{13} \\
 -c_{12} s_{13} s_{23} e^{i \delta_{CP}}-c_{23} s_{12} & c_{12} c_{23}-e^{i \delta_{CP}} s_{12} s_{13} s_{23} & c_{13} s_{23} \\
 s_{12} s_{23}-c_{12} c_{23} e^{i \delta_{CP}} s_{13} & -c_{23} s_{12} s_{13} e^{i \delta_{CP}}-c_{12} s_{23} & c_{13} c_{23} \\
\end{array}
\right) \times \mathrm{P},\,\,\, \label{Ulepg}
\eea
we can parameterize the  solar neutrino mixing angle $\theta_{12}$, the Dirac CP phase $\delta_{CP}$ and the model parameters $\theta, \Psi, \eta_{1,2}$ in terms of the other two neutrino mixing angles $\theta_{13}$ and $\theta_{23}$ as follows:%\footnote{Here, we have defined $c_{kl}=\cos \theta_{kl}, s_{kl}=\sin \theta_{kl}, s'_{kl}=\sin 2\theta_{kl}, c'_{kl}=\cos 2\theta_{kl}\, (kl=12, 13, 23)$, $s'_\theta=\sin 2\theta, c'_\theta=\cos 2\theta$ and $c_\Psi=\cos\Psi$. }
  \bea
    &&s^2_{12} c^2_{13} = \frac{1}{3}  \hspace{0.2cm}\mbox{for both NH and IH}, \label{s12c13}\\
    &&\cos\theta     =\left\{
\begin{array}{l}
-\sqrt{\frac{1}{2} - \frac{\sqrt{3}}{2}\sqrt{c^4_{13} \sin^2 2\theta_{23}-\cos^2 2\theta_{13}}} \hspace{0.3cm}\mbox{for NH},    \\
\hspace{0.3 cm} \sqrt{\frac{1}{2} + \frac{\sqrt{3}}{2}\sqrt{c^4_{13} \sin^2 2\theta_{23}-\cos^2 2\theta_{13}}} \hspace{0.25cm}\,\mbox{for IH},
\end{array}%
\right. \label{costhetav}\\
&&\cos\psi=  \frac{1-3 s_{13}^2}{2\sqrt{1-\frac{3}{4}\left(\sin^2 2\theta_{13}+ c_{13}^4 \sin^2 2\theta_{23}\right)}} \hspace{0.2cm}\mbox{for  both NH and IH,} \label{Psi}\\
&& \sin\delta_{CP}=-\frac{2\sqrt{c^4_{13} \sin^2 2\theta_{23}-\cos^2 2\theta_{13}}}{s_{12} \sin 2\theta_{13} \sin 2\theta_{23}\sqrt{6-9 s^2_{13}}} \hspace{0.2cm}\mbox{for  both NH and IH}, \label{sintheta}\\
&&\eta_1= \left\{
\begin{array}{l}
-i \log \left(\frac{c_\theta -s_\theta e^{i \psi}}{\sqrt{3} c_{12} c_{13}}\right) \hspace{0.2cm}\mbox{for  NH},    \\
-i \log \left(\frac{s_\theta + c_\theta e^{i \psi}}{\sqrt{3} c_{12} c_{13}}\right) \hspace{0.175cm}\,\mbox{for  IH},
\end{array}%
\right. \\
&& \eta_2=\left\{
\begin{array}{l}
\delta_{CP}-i \log \left(\frac{c_\theta +s_\theta e^{-i \psi}}{\sqrt{3} s_{13}}\right) \hspace{0.2cm}\mbox{for NH},    \\
\delta_{CP}-i \log \left(\frac{s_\theta -c_\theta e^{-i \psi}}{\sqrt{3} s_{13}}\right) \hspace{0.2cm} \mbox{for IH},
\end{array}%
\right.\label{eta2v}
\eea
Next, from eqs. (\ref{m123sq}) and (\ref{theta}) we can express the parameters $X_0, Y_0, A^2_0, C^2_0, D^2_0$, two neutrino masses ($m_{1,3}$) and the sum of neutrino masses ($\sum m_i$) in terms of the second neutrino mass $m_2\equiv B_0$ and two squared mass differences $\Delta m^2_{21}$, $\Delta m^2_{31}$ for NH ($\Delta m^2_{21}$, $\Delta m^2_{32}$ for IH) as follows
\bea
&& X_0=\left\{
\begin{array}{l}
\sqrt{B_0^2-\Delta m^2_{21}+\frac{\Delta m^2_{31}}{2}} \hspace{0.25cm}\mbox{for  NH},    \\
\frac{1}{\sqrt{2}}\sqrt{B_0^2+\frac{\Delta m^2_{32}-\Delta m^2_{21}}{2}} \hspace{0.1cm}\,\mbox{for IH},
\end{array}%
\right. \,\, Y_0=\left\{
\begin{array}{l}
\sqrt{\frac{\Delta m^2_{31}}{2}} \hspace{1.5cm}\mbox{for NH},    \\
\sqrt{-\frac{\Delta m^2_{21}+\Delta m^2_{32}}{2}} \hspace{0.1cm}\,\mbox{for IH},
\end{array}%
\right. \label{X0Y0}\\
&&A^2_0=\left\{
\begin{array}{l}
B_0^2 -\Delta m^2_{21}+ \Delta m^2_{31}\sin^2\theta \hspace{1.25cm}\mbox{for NH},    \\
B_0^2 - \Delta m^2_{21} \cos^2\theta + \Delta m^2_{32}\sin^2\theta \hspace{0.25cm}\mbox{for IH},
\end{array}%
\right. \label{A0sq}\\
&&C^2_0=\left\{
\begin{array}{l}
B_0^2 -\Delta m^2_{21}+ \Delta m^2_{31}\cos^2\theta\hspace{1.2cm}\mbox{for NH},    \\
B_0^2 -\Delta m^2_{21} \sin^2\theta+ \Delta m^2_{32}\cos^2\theta\hspace{0.25cm}\mbox{for IH},
\end{array}%
\right. \label{C0sq}\\
&&D^2_0=\left\{
\begin{array}{l}
\Delta m^2_{31}\sin\theta\cos\theta \hspace{2.3cm}\mbox{for NH},    \\
-(\Delta m^2_{21}+\Delta m^2_{32})\sin\theta\cos\theta \hspace{0.25cm}\mbox{for  IH},
\end{array}%
\right. \label{D0sq}\\
&&\left\{
\begin{array}{l}
m_1=\sqrt{B^2_0-\Delta m^2_{21}},\,\, m_2=B_0, \,\, m_3=\sqrt{B^2_0-\Delta m^2_{21}+\Delta m^2_{31}}\hspace{0.3cm}\mbox{for NH},    \\
m_1=\sqrt{B^2_0-\Delta m^2_{21}},\,\, m_2=B_0, \,\, m_3=\sqrt{B^2_0+\Delta m^2_{32}}\hspace{1.75cm}\mbox{for IH},
\end{array}%
\right. \label{mi}\\
&&\sum m_i=\left\{
\begin{array}{l}
B_0+\sqrt{B^2_0-\Delta m^2_{21}}+\sqrt{B^2_0-\Delta m^2_{21}+\Delta m^2_{31}}\hspace{0.3cm}\mbox{for NH},    \\
B_0+\sqrt{B^2_0-\Delta m^2_{21}}+\sqrt{B^2_0+\Delta m^2_{32}}\hspace{1.75cm}\mbox{for IH}.
\end{array}%
\right. \label{summi}
\eea
Expression (\ref{s12c13}) shows the possible range %\footnote{Here, numbers are displayed with 4 significant digits to the right of the decimal point.}
 of $s^2_{12}$ with $s^2_{12}\in (0.3402, 0.3416)$ %\, ($\theta_{12} \in (35.68, 35.76)^\circ$)
in the case of $s^2_{13}\in (0.02032, 0.02410)$ % \, ($\theta_{13} \in (8.20, 8.39)^\circ$)
 for  NH \cite{Esteban2021} and $s^2_{12}\in (0.3403, 0.3416)$ %\, ($\theta_{12} \in (35.69,35.77 )^\circ$)
  in  the case of $s^2_{13}\in (0.0.02052, 0.02428)$ %\, ($\theta_{13} \in (8.24, 8.96)^\circ$)
  for IH \cite{Esteban2021}. Futhermore, at $3\sigma$ range, $s^2_{23}\in (0.415,\, 0.616)$ and $s^2_{13}\in (0.02032, 0.02410)$ for NH while $s^2_{23}\in (0.419,\, 0.617)$ and $s^2_{13}\in (0.0.02052, 0.02428)$ for IH \cite{Esteban2021}. Expressions (\ref{costhetav})--(\ref{eta2v}) yield the possible ranges of $\cos\theta, \cos\psi, \sin\delta_{CP}$ and $\eta_{1,2}$:
\bea
&&\cos\theta \in \left\{
\begin{array}{l}
(-0.640, -0.570) \hspace{0.2cm} \mbox{for  NH},    \\
(0.770, 0.820)  \hspace{0.72cm}\,\mbox{for IH},
\end{array}%
\right.,\mathrm{i.e.},\, \theta \,(^\circ) \in \left\{
\begin{array}{l}
(124.80, 129.80) \hspace{0.15cm} \mbox{for  NH},    \\
(34.92, 39.65)  \hspace{0.5cm}\,\mbox{for IH},
\end{array}%
\right.  \label{costhetar}\\
&&\cos\psi \in\left\{
\begin{array}{l}
(0.930, 0.990) \hspace{0.2cm} \mbox{for  NH},    \\
(0.92, 0.99)  \hspace{0.5cm}\,\mbox{for  IH},
\end{array}%
\right.\mathrm{, i.e.,} \, \psi\, (^\circ) \in \left\{
\begin{array}{l}
(8.11, 21.57) \hspace{0.2cm} \mbox{for  NH},    \\
(8.11, 23.07)  \hspace{0.1cm}\,\mbox{for IH},
\end{array}%
\right.  \label{cosvarr}\\
&&\eta_{1}\, (^\circ)\in\left\{
\begin{array}{l}
(182.40, 189.30) \hspace{0.2cm} \mbox{for  NH},    \\
(1.146, 9.167) \hspace{0.5cm}\,\mbox{for  IH},
\end{array}%
\right.\,\,\,\,\,  \eta_{2} \, (^\circ)\in\left\{
\begin{array}{l}
(260.30, 268.30)\hspace{0.2cm} \mbox{for  NH},    \\
(80.21, 88.24)\hspace{0.5 cm}\,\mbox{for  IH},
\end{array}%
\right. \label{bet31f}\\
&&\delta_{CP}\, (^\circ)\in\left\{
\begin{array}{l}
(279.80, 325.60)\,\hspace{0.15cm} \mbox{for  NH},    \\
(295.80, 330.0)\, \hspace{0.25 cm}\,\mbox{for  IH}.
\end{array}%
\right. \label{bet31f}
\eea
Turning to the neutrino sector, eqs. (\ref{m123sq}), (\ref{costhetav}) and (\ref{X0Y0})$-$(\ref{summi}) show that at the best-fit values of the differences of the squared neutrino masses, $X_0$, $m_{1,3}$ and $\sum m_i$ depend on only one parameter $B_{0}\equiv m_2$; $D_0$ depends on only one parameter $\theta$ while $A_0$ and $C_0$ depend on two parameters $B_0$ and $\theta$.
With the aid of (\ref{costhetar}), i.e., $\cos\theta \in (-0.64, -0.57)$ for NH and $\cos\theta \in (0.77, 0.82)$ for IH, we get
 \bea
D^2_{0}\in\left\{
\begin{array}{l}
(1.18,\, 1.23)\times 10^{-3}\, \mathrm{eV}^2\hspace{0.2cm} \mbox{for  NH},    \\
(1.14, 1.19)\times 10^{-3}\,  \mathrm{eV}^2\hspace{0.2 cm}\,\mbox{for  IH}.
\end{array}%
\right. \label{D0region}
\eea
Furthermore, the cosmological observations provide the constraints on neutrino masses $\sum m_i  < 120\, \mathrm{meV}$ (for NH) and $\sum m_i  < 150\, \mathrm{meV}$ (for IH) \cite{Salas2021, Aghanim20}. In order to achieve these constraints, in present study, we consider $B_0\equiv m_2 \in \left(1.0, 3.0\right)\times 10^{-2}\, \mathrm{eV}$ (for NH) and $B_0 \in \left(5.0, 6.0\right)\times 10^{-2}\, \mathrm{eV}$ (for IH). At the best-fit values of the differences of the squared neutrino masses, $\Delta m^2_{21}=7.42\times 10^{-5} \, \mathrm{eV}^2$, $\Delta m^2_{31}=2.517\times 10^{-3} \, \mathrm{eV}^2$ for NH and $\Delta m^2_{32}=-2.498\times 10^{-3} \, \mathrm{eV}^2$ for IH, we get:
\bea
&&Y_{0}=\left\{
\begin{array}{l}
3.548\times 10^{-2}\, \mathrm{eV}\hspace{0.35cm} \mbox{for  NH},    \\
3.481\times 10^{-2}\,  \mathrm{eV}^2\hspace{0.15cm}\mbox{for  IH},
\end{array}%
\right. \label{X0region}\hspace{1.0 cm} X_{0}\in\left\{
\begin{array}{l}
(3.58,\, 4.56)10^{-2}\, \mathrm{eV}\hspace{0.25cm} \mbox{for  NH},    \\
(3.60, 4.80)10^{-2}\,  \mathrm{eV}^2\hspace{0.2cm}\mbox{for  IH},
\end{array}%
\right. \label{X0region}\\
&&A_{0}^2\in\left\{
\begin{array}{l}
(1.60,\, 2.40)\times 10^{-3}\, \mathrm{eV}\hspace{0.25cm} \mbox{for  NH}, \\
(1.60, 2.60)\times 10^{-3}\,  \mathrm{eV}^2\hspace{0.2cm}\,\mbox{for  IH},
\end{array}%
\right. \, C_{0}^2\in\left\{
\begin{array}{l}
(1.0,\, 1.80)\times 10^{-3}\, \mathrm{eV}\hspace{0.35cm} \mbox{for  NH},    \\
(0.80, 2.00)\times 10^{-3}\,  \mathrm{eV}^2\hspace{0.15cm}\mbox{for  IH},
\end{array}%
\right. \label{C0region}\\
&&m_{1}\in\left\{
\begin{array}{l}
(0.5,\, 2.88\times 10^{-2})\, \mathrm{eV}\hspace{0.35cm} \mbox{for  NH},  \\
(5.00, 5.80)\times 10^{-2}\,  \mathrm{eV}^2\hspace{0.15cm}\,\mbox{for  IH},
\end{array}%
\right.
%\\&&m_{2}\in\left\{
%\begin{array}{l}
%(1.0,\, 3.0)\times 10^{-2}\, \mathrm{eV}\hspace{0.25cm} \mbox{\Red{for  NH}},    \\
%(5.00, 6.00)\times 10^{-2}\,  \mathrm{eV}^2\hspace{0.2cm}\,\mbox{\Red{for  IH}}.
%\end{array}%
%\right. \label{m2region}\\
%&&
\, m_{3}\in\left\{
\begin{array}{l}
(5.04,\, 5.78)\times 10^{-2}\, \mathrm{eV}\hspace{0.32cm} \mbox{for  NH},    \\
(0.50,\, 3.00)\times 10^{-2}\,  \mathrm{eV}^2\hspace{0.15cm} \mbox{for  IH},
\end{array}%
\right. \label{m3region}\hs\, \\
&&\sum m_{i}\in\left\{
\begin{array}{l}
(0.065,\, 0.1167)\, \mathrm{eV}\hspace{0.2cm} \mbox{for  NH},    \\
(0.11, 0.150)\,  \mathrm{eV}^2\hspace{0.5 cm}\mbox{for  IH}.
\end{array}%
\right. \label{summiregion}
\eea
%\section{\label{effect}Effective neutrino mass parameters}
%From Eqs. (\ref{m1m2m3}) and (\ref{Ulep}), t
At this stage, the effective neutrino masses governing the neutrinoless double beta decay, $\langle m_{ee}\rangle = \left| \sum^3_{i=1} U_{ei}^2 m_i \right|$, and beta decay, $m_{\beta }= \left(\sum^3_{i=1} \left|U_{ei}\right|^2 m_i^2\right)^{1/2}$, are determined as
\bea
\langle m_{ee}\rangle\in \left\{
\begin{array}{l}
(0.75,\, 2.75)\times 10^{-2}\, \mbox{eV}\hspace{0.2cm}  \mbox{for NH,} \\
(4.8,\, 5.8)\times 10^{-2}\, \mbox{eV}\hspace{0.55cm} \mbox{for IH,} %
\end{array}%
\right.   \,
m_{\beta}\in \left\{
\begin{array}{l}
(1.25,\, 3.0)\times 10^{-2}\,\mbox{eV}\hspace{0.15cm} \mbox{for  NH,} \\
(5.0,\, 5.8)\times 10^{-2}\, \mbox{eV}\hspace{0.35cm}  \mbox{for IH,}
\end{array}%
\right.   \label{membetvalues}\,
\eea
provided that $B_0\in (0.01, 0.03)$, $s^2_{23}\in (0.415, 0.616)$ and $s^2_{13}=0.02219$ for NH \cite{Esteban2021} and $B_0\in (0.05, 0.06)$, $s^2_{23}\in (0.419, 0.617)$ and $s^2_{13}=0.02238$ for IH \cite{Esteban2021}.

%which are below all the upper limits arising from recent $0\nu \beta \beta $ decay experiments taken from KamLAND-Zen \cite{KamLAND16} $\langle m_{ee} \rangle <0.061 \div 0.165\, \mathrm{eV}$, GERDA \cite{GERDA19} $\langle m_{ee} \rangle < 0.104\div 0.228\, \mathrm{eV}$, and CUORE \cite{CUORE20} $\langle m_{ee} \rangle < 0.075 \div 0.35 \,\mathrm{eV}$.

 The above analysis shows that the considered model can accommodate \emph{both normal and inverted mass hierarchies} which is completely different from the work in Ref.\cite{A4Datta} where \emph{only the normal neutrino mass hierarchy is allowed}. Furthermore, the leptonic mixing matrix in current work is the same as that of Ref.\cite{A4Datta}, however, the expressions of neutrino masses are different from each other\footnote{We check that, with $U_{13}$ is given in Eq. (2.16) of Ref. \cite{A4Datta} for arbitrary $\psi$, Eq. (2.15) of Ref. \cite{A4Datta}, $U^T_{13} Y^\nu Y^{\nu T} U_{13} =\mathrm{diag} (\lambda_1, \lambda_2, \lambda_3)$, is not true. This equation %(2.15) of Ref. \cite{A4Datta}
is just satisfied if $\psi=0$.
Furthermore, substituting Eq. (2.15) in Eq. (2.13) to obtain Eq. (2.21) occurs only if and only if $\psi=0$ to ensure that $U_{13}$ is an orthogonal matrix. %, i.e., $U_{13} U_{13}^T=1$.
Unfortunately, for arbitrary $\psi$, $U_{13}$ in Eq. (2.16) of Ref. \cite{A4Datta} is not an orthogonal matrix since $U_{13} U_{13}^T\neq 1$\, (In fact, $U_{13}$ in Eq. (2.16) of Ref. \cite{A4Datta} is a unitary matrix, $U_{13} U_{13}^\+ = 1$, and $U_{13}$ is a orthogonal matrix if and only if $\psi=0$).}.
%%%%%%%%%%%%%%%%%%%%%%%%%%%%%%%%%%%%%%%%%%%%%%%%%%%%%%%%%%%%%%%%%%%%%%%%
%\section{Conclusion}

\section{Conclusion}

We have shown that the invariant Yukawa term $\frac{z^\nu}{2\Lambda} \left[\left(\overline{N^c_R} N_R\right)_{\mathbf{3_s}} \left(\varphi \Psi^* +\varphi^* \Psi\right)_{\mathbf{3}}\right]_{\mathbf{1}}$ must be added to the Yukawa terms of ref. \cite{A4Datta}. When this term is included, the obtained results are significantly different from the ones presented in ref. \cite{A4Datta}. Namely, the structure of Dirac neutrino mass matrix $M_D$ is the same as that of ref. \cite{A4Datta}, however, the structure of Majorana mass matrix $M_R$ is completely different from each other. We also shown that the considered model can accommodate both normal and inverted mass hierarchies which is completely different from the work in ref.\cite{A4Datta} where only the normal neutrino mass hierarchy is allowed. In addition to the Dirac CP phase, Majorana phases, which has been ignored in ref.\cite{A4Datta}, are also predicted to be $\eta_{1}\in (182.40, 189.30)^\circ, \eta_2 \in (260.30, 268.30)^\circ$ for the normal hierarchy and $\eta_1 \in (1.146, 9.167)^\circ, \eta_2\in (80.21, 88.24)^\circ$ for the inverted hierarchy.

%\acknowledgments

%This is the most common positions for acknowledgments. A macro is available to maintain the same layout and spelling of the heading.

%\paragraph{Note added.} This is also a good position for notes added after the paper has been written.


\begin{thebibliography}{99}

\bibitem{A4Datta} A. Datta, B. Karmakar, A. Sil, \emph{Flavored leptogenesis and neutrino mass with $A_4$ symmetry}, J. High Energ. Phys. 2021 (2021) 51.% https://doi.org/10.1007/JHEP12(2021)051.
\bibitem{A4Dattaconf} A. Datta, B. Karmakar, A. Sil, \emph{A Minimal Flavor Model for Neutrino Mass and Leptogenesis},
Acta Phys. Pol. B Proc. Suppl. 15, 2-A13 (2022).
\bibitem{Esteban2021} I. Esteban, M.C. Gonzalez-Garcia, M. Maltoni, T. Schwetz and A. Zhou, \emph{The fate of hints:
updated global analysis of three-flavor neutrino oscillations}, J. High Energ. Phys. 2020, 178 (2020). %https://doi.org/10.1007/JHEP09(2020)178. %arXiv:2007.14792 [hep-ph].
\bibitem{Salas2021} P. F. de Salas \emph{et al.}, \emph{2020 Global reassessment of the neutrino oscillation picture}, J. High Energ. Phys. 2021, 71 (2021). %https://doi.org/10.1007/JHEP02(2021)071. %, arXiv: 2006.11237 [hep-ph].
\bibitem{Aghanim20} N. Aghanim \emph{et al.} [Planck Collaboration], \emph{Planck 2018 results. VI. Cosmological parameters},
Astron. Astrophys. 641 (2020) A6. %https://doi.org/10.1051/0004-6361/201833910.%, arXiv: 1807.06209 [astro-ph.CO].

\end{thebibliography}
\end{document}